# Design of a Self-powered Smart Mask for COVID-19

Barnali Ghatak[†], Sanjoy Banerjee[§], Sk Babar Ali[‡], Rajib Bandyopadhyay[*,†,#]

Nityananda Das[*,⊥], Dipankar Mandal[*, ‖], Bipan Tudu[*,†]

[†]Department of Instrumentation and Electronics Engineering, Jadavpur University, Kolkata 700106, India

[§]Department of Applied Electronics and Instrumentation Engineering, Future Institute of Engineering and Management, Kolkata 700150, India

[‡]Department of Electronics and Communication Engineering, Future Institute of Engineering and Management, Kolkata 700150, India

[#]Laboratory of Artificial Sensory Systems, ITMO University, Saint Petersburg, 191002, Russia

[⊥]Department of Physics, Jagannath Kishore College, Purulia 723101, West Bengal, India

[‖]Institute of Nano Science and Technology (INST), Habitat Centre, Phase 10, Sector 64, Mohali 160062, India

**Corresponding authors:** RB: bandyopadhyay.rajib@gmail.com; ND: ndas228@yahoo.com; DM: dmandal@inst.ac.in; BT: bipantudu@gmail.com

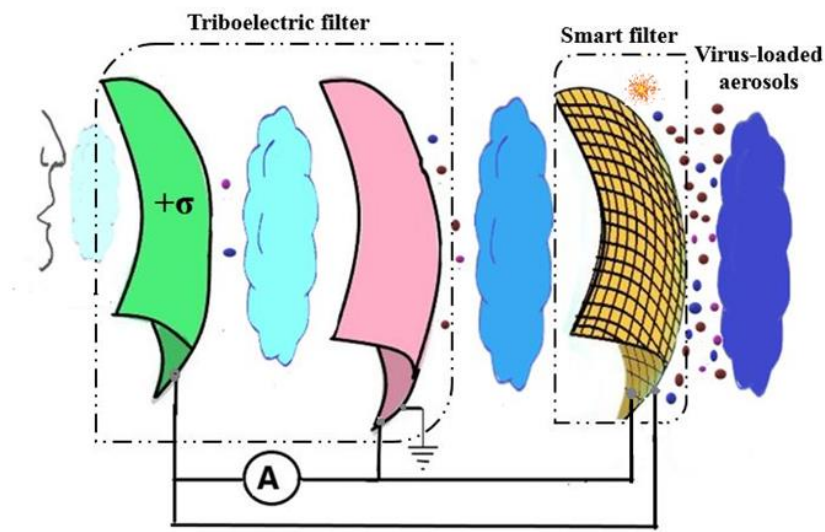

# ABSTRACT

Usage of a face mask has become mandatory in many countries after the outbreak of SARS-CoV-2, and its usefulness in combating the pandemic is a proven fact. There have been many advancements in the design of a face mask and the present treatise describes a face mask in which a simple textile triboelectric nanogenerator (TENG) serves the purpose of filtration of SARS-CoV-2. The proposed mask is designed with multilayer protection sheets, in which the first two layers act as triboelectric (TE) filter and the outer one is a smart filter. The conjugated effect of contact electrification, and electrostatic induction of the proposed smart mask are effective in inactivating the span of virus-ladden aerosols in a bidirectional way. Five pairs of triboseries fabrics i.e. nylon - polyester, cotton - polyester, poly(methyl methacrylate) - PVDF, lylon - PVDF and polypropylene - polyester have been optimized in this study in terms of their effective tribo-electric charge densities as 83.13, 211.48, 38.62, 69 and 74.25 $nC/m^2$, respectively. This smart mask can be used by a wide range of people because of its simple mechanism, self-driven (harvesting mechanical energy from daily activities, e.g. breathing, talking, or other facial movements functionalities, and effective filtration efficiency and thus, it is expected to be potentially beneficial to slow down the devastating impact of COVID-19.

KEYWORDS: *Textile masks,TENG, contact electrification, SARS-CoV-2, self-powered mask, COVID-19*

## 1. Introduction

While the ongoing outbreak of 2019-2020 severe acute respiratory syndrome coronavirus 2 (SARS-CoV-2, previously known as 2019-nCoV) gains significant prevalence worldwide, scavenging a curative remedy has become the main thrust to all scientists across the globe. Alike personal protective equipment (PPE), usage of homemade do-it-yourself (DIY) cloth mask has attracted global attention to combat novel coronavirus. Though the performance of various fabrics used in such a homemade mask is still under observation, their response is likely to be anticipated for combatting severe respiratory issues caused by the novel pandemics. In regards to the present scenario, the recent report by Konda et al. illustrated the filtration efficiency of differently combined fabrics depends on the particle size of aerosol, and its found to be promising for the wide bracketed size (10 nm- 6µm) particles. Their study unveils aerosol filter made by the combination of chiffon and silk that are significantly efficient.[1] Therefore, a hybrid combination of non-woven fabrics having different threads count per unit shows effective filtration efficiency fornanosize particles. Improper and loose-fitting might prevent the working of that mask as the filtration is based on the principle of static electricity.

Kutter et al. have well demonstrated the variation in the size of respiratory droplet, which are well known as aerosols (<5 µm) and droplets (>5 µm).[2] Among these, water droplets and liquefied gas droplets come into play as a medium on which SARS-CoV-2 can traverse over a long distance when the size is relatively small,. Otherwise, larger droplets can settle down easily without traveling a large distance.

Based on the available knowledge of various aspects of infection spread by SARS-CoV-2 virus, the initialspreading is likely to be spread by direct contact, or by providing a transmission path to land the large virus-containing droplets,[3] which are found to remain stable for more than 24 hrs.[4]

COVID positive patients can also passively spread viruses to the people surrounding them. As per the underlying precaution made by WHO 2020a, wearing a face mask is mandatory to combat COVID-19.[5] Another route of transmission might be of either only in passing or by traversing virus contained aerosols. So in that case, the liquefied droplets start to evaporate immediately after the expiry of droplets and, some of the smaller sized droplets transported by air discharge field rather than gravitation. Such small virus-loaded droplets get enough liberation to traverse in the air several meters from their route.[6] Following five basic steps of aerosol filtration, the electrostatic attraction takes predominant role alike gravity sedimentation, inertial impaction, interception, and diffusion.[7-8] All these mechanisms are solely size-dependent, i.e. gravitational forces come into play for larger sized droplets ($> 1\mu m$), whereas sedimentation and impaction are only applicable for medium-sized particles ($> 1\mu m$ to $10\mu m$). Lesser the size of the particles (100 nm- $11\mu m$), higher will be the tendency of those particles to get diffused by mechanical interception, and Brownian motion. Interestingly, nanosized particles can easily slide between the opening network of non-woven fibers.[9] In such cases, electrostatic attraction (EA) takes significance to bind and cling such particles to the fibers. The newly invented principle of fusing EA with nanogenerators capable of generating triboelectric charges from the wasted mechanical energy from our daily activities (respiration, talking, or any other facial movement) can be utilized favorably in mask design,[10] which has been conceptualized in the present study.

In this context, we present the design and simulation of a novel self-powered triboelectric nanogenerator (TENG) fused smart mask, in which the viruses are killed in the electric field and the wearer can combat the deadly novel coronavirus. Recent findings on the use of cloth mask during the outbreak of Influenza in 2009 appeared to be inadequate to conclude the selection of fabrics having significant filtration efficiency.[11-12] Though the recent study by Konda et al. have envisaged that the thread counts of respective non-woven fiber play a key role in electrostatic filtration, but does not guarantee for the case of inhaling viral particles coming from the environment.

**2. Proposed design of face mask**

The proposed face mask is composed of multilayers, out of which inner and middle layers comprise of tribo-series materials (TSM). The prototype design of the self-powered smart mask is shown in **Figure 1(a)**. Since TENGs exhibit quadratic relation between power density, and triboelectric charge density, increasing the tribo- charges is a challenging issue.[13-16] That can be done using structural optimization of different tribo-series materials, surface modification, etc.[17] Interestingly, nucleocapsid protein crowned SARS-CoV-2 possesses surface electrostatic potential characteristics, that reinforced the present design concept.[18] Keeping all these in mind, five combinations of readily available non-woven TSM have been chosen, each with a pair of positive TSM and negative TSM followed by a self-powered smart layer. As a good energy harvester, vocal energy exerted during talking, or any other lip activities have been utilized as the prime source of power to induce static electricity in between the inner and middle layers of the proposed mask. The detailed characteristics of the five pairs of TSM are illustrated in the next section. The self-powered induced potential allows the thin metal framing of these layers to transfer charges between them. The layer is capable of acquiringthe static charges so that any viral particle having surface charge can easily be inactivated in the outer layer.[19] The activated outer or smart layer can adhere electrically the charged viral particles coming from the close vicinity of the wearer. The novelty of this design lies in the self-activation of the smart layer through the vocal activities of the wearer and has provision to deal with aerosols as well as droplets (gaseous, liquid). The essence of the proposed multilayer self-powered smart mask is low-cost, comfortable texture,and wide accessibility thatassures no internal respiratory problem (breathing problem) to the wearer.

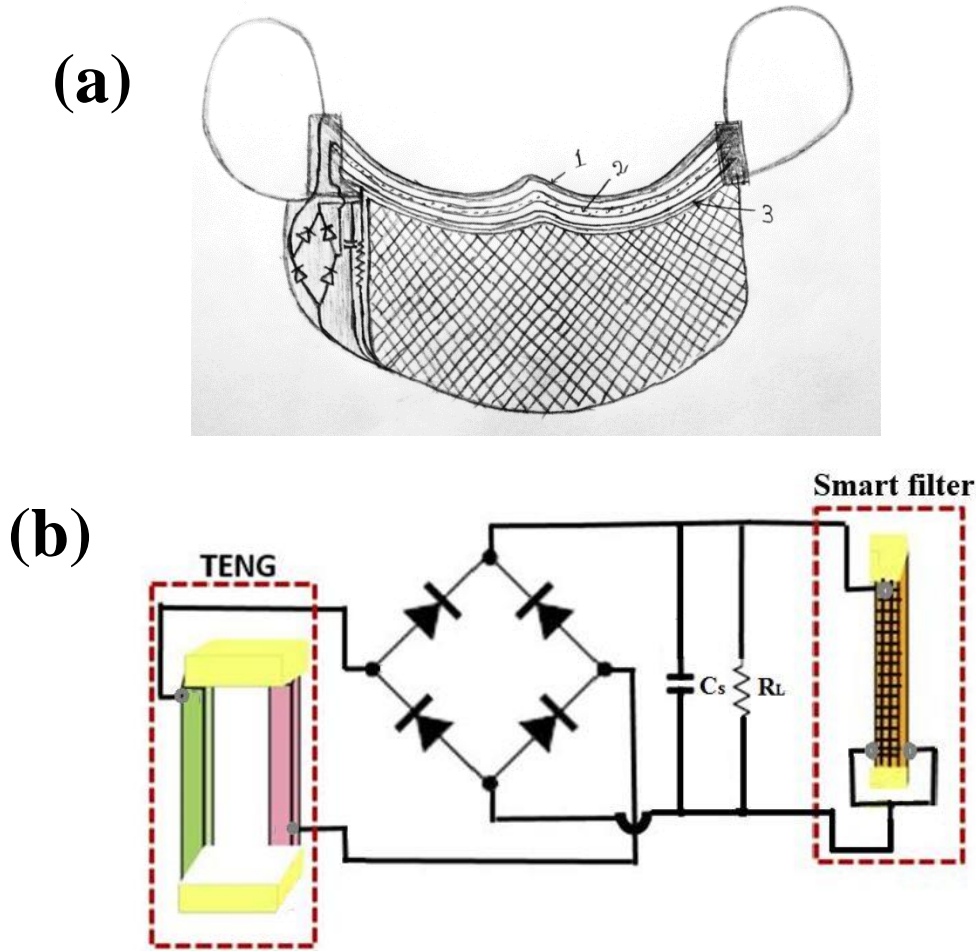

**Figure 1. (a) The proposed self-powered smart mask (1- inner layer, 2-middle layer, 3- smart layer). (b) Schematic representation of working mechanism TENG- based smart mask. The proposed mask can filtrate virus contained aerosols both during breath in and breath out condition. The outer smart layer gets shorted due to the charge generated by TENG. The short-circuited net mesh can fetch viral particles having surface charge within its filter layer. Consequently, the particle becomes inactive through electrocution in the outer layer. The**

**tribo-layers provide extra protection to the mask wearer. Here C$_s$ indicates storage capacitor, and R$_L$ is the load resistor.**

In this paper, the simple mechanism of auto harvesting energy, contact electrification, and electrostatic induction enabled TENG have been utilized. The comparative study of reported design face mask is explained in **Table 2**. The corresponding features of triboseries fabrics belonging to the positive and negative series of the tribo-series is described in **Table 1** . To boost up the output charge of TENG, combination of nylon - polyester, cotton - polyester, PMMA - PVDF, lylon - PVDF and polypropylene - polyester have been studied. Based on this, the electrostatic simulation has been carried out.

## 3. Theoretical simulation of the self-powered smart mask

The principle of TENGs isfusion of contact electrification and electrostatic induction. Static polarized charges are mainly induced through contact electrification which further triggers the energy conversation from self-harvested vocal energy to electrical energy through electrostatic induction.

Initially, the separation distance ($d$) between two tribo-pairs can be varied according to the activity of the liops of the wearer. The mechanical force exerted by the lip activity makes them come in contact with each other. The material in each tribo-pairs share opposite tribo-charges as aresult of contact electrification. The insulated construction of each layer (**Figure 1 b**), confirms that the charges can only transfer between tribo-layers (TL) through external circuits. If '$+\sigma$' denotes the transferred charges between TLs, one TL will acquire transferred charge $-\sigma$ and the other TL will have the transferred charge of $+\sigma$. In Figure 2A, the separation distance between the TLs is '$d$' with an area of '$S$'. We consider the size of the virus-loaded particle to be 1µm. The electrical potential difference between the TLs of TENG is supposed to contribute two major components.

The induced polarized TE generates charges and voltage, which is a function of '$d$'. Considering the induced voltage to be 1V, the outgoing viral particles will be under very high electric field (E), i.e. $10^6$ V/m, as calculated using equation 1.

$$E = \frac{V}{d} \qquad (1)$$

Therefore, the generated higher electric field is capable enough to make a contact between these TLs, since the air discharge field is $3 \times 10^6$ V/m.

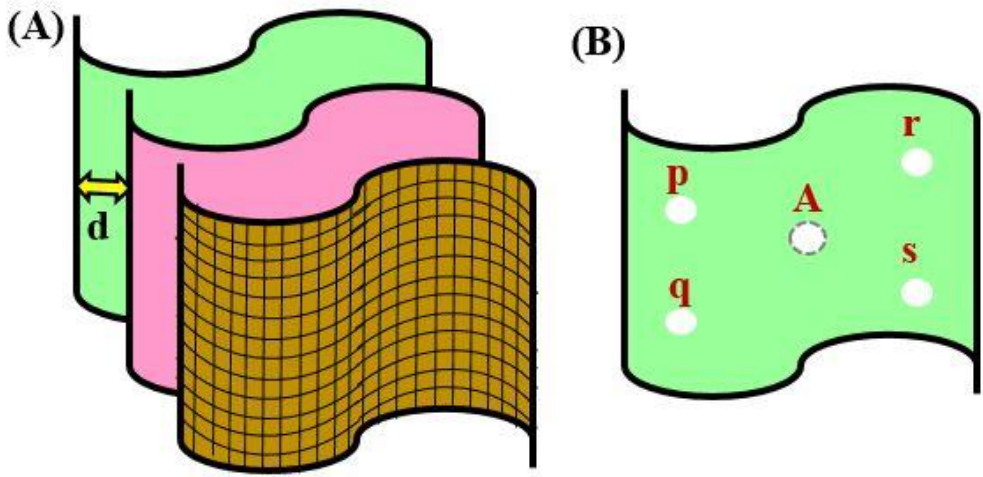

**Figure 2. (A) Alignment of three-layer of the proposed smart mask. (B) Considering four holes in the inner layer with another hole in the opposite side.**

A deep insight into the converter circuit illustrates that the voltage output of the TENG gets rectified using the bridge rectifier. The charging capacitor ($C_s$) stores the accumulated charge into it for future use. It removes the unwanted AC components of the output signal. Thus, a pure DC is obtained across the load resistor $R_L$. The smart filter of the proposed mask is connected across the $R_L$. Therefore the induced tribo charges can be discharged for $R_L C_s$ time (sec). This finding provides the wearer extra protection while the wearer are at rest.

The other perspective illuminates that the viral respiratory droplets are of conducting nature, hence their resistance will be smaller (1kΩ). Also, 1mA current through the human body is sufficient enough to inactivate the surface charge of the virus. Therefore, the two TLs of 1µm apart with self-generated voltage in the mask is around 1 V, that indicates the aerosolized droplet can easily be burnt through short-circuiting in the third smart layer. If the test droplet does not contain critical amount of surface charge, then also water particles make the layer shorted.

It's imperative to measure the safety of the wearer concerning the generated heat energy,illustrate this, four holes (p, q, r, s) are considered in the front side, whereas the hole in the opposite side the respected layer is depicted by 'A' (**Figure 2B**). The tendency of the incoming droplets entering through any of the holes (p, q, r, s) to emanate through the hole 'A' in the opposite side of the layer. During the passage from TLs to the smart layer, the droplet has to adhere to the mesh of the smart layer that has already been energized highly. Therefore, the high field induced in the smart layer capable of electrocuting the outgoing droplet by heat burning. The produced heat would be in the order of $1fJ/\mu m^3/°k$ (considering the air specific density ~ $1kJ/m^3/°k$). This phenomenon further assures the heat energy produced to electrocute the outgoing particles would not create much heat to initiate any breathing problems to the wearer.

**4. Power management of self-powered smart mask**

It has been illustrated in **Table 1.** based on this mask design based on static electricity, triboelectricity, etc. Analyzing the electric characteristics, and material properties of TE materials can conclusively give a deep insight into the material dependent voltage, current, developed power in the TENG. Considering the the normal respiration rate for an adult 12-20 breaths/ min. at rest. In order to calculate minimum available energy, the frequency bandwidth (fmax-fmin) of the expiration phase is 0.2-0.33 Hz. Since, one complete breathing cycle requires 5 sec, inspiration and expiration ratio (I:E) would be 1:1 (2.5 sec each) during active phase of the wearer. Therefore, the effect of a variation in inspiration and expiration times subjected to a pressure difference of 1kPa (**Supplementary Information, Note S1**).

Since the novelty of the proposed mask lies in the smart layer, the utilization of the developed charged between TLs should be sufficient to allow electrocution in the smart layer.

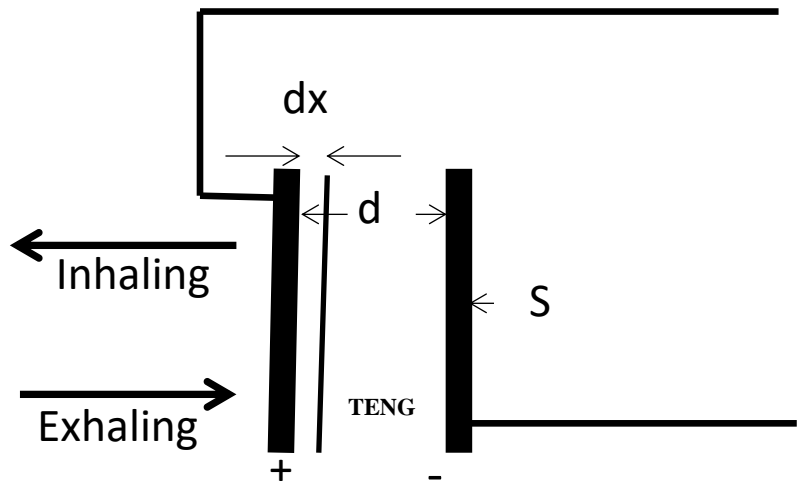

**Figure 3. Arrangement and alignment of TLs during inspiration and expiration. The figure indicating the constant separation distance 'd' between the two TLs with area of each layer 'S' and the relative separation distance during breath in and breath out is denoted by 'dx'.**

In order to analyze the charge density of two TLs based TENG, average energy available from breathing can be calculated as $5 \times 10^5 \times k$ J (considering $S \sim$ 5 mm), as shown in **Figure 3**. Available charge during each breathing cycle supply charge $(d_q)$ of $5 \times 10^5 k \times n$C/ breathing ('$n$' denotes the effective triboelectric charge density of respective pair, TECD). Therefore, the important parameter '$k$' signifies the variation of energy required for charging which depends on the internal properties of TSMs under study. **Table 1** is illustrating the TECD of nylon - polyester, cotton - polyester, PMMA - PVDF, nylon - PVDF, and polypropylene - polyester. It can be understood from **Table 1** that the pair of cotton- polyester TLs likely to be the suitable tribo-pairs with highest effective TECD. The proposed design of the proposed mask thus optimized to the pair of cloths such as a cotton and non-woven polyester fabrics. Besides, the relative charge available during active phase of the wearer can be calculated as $500 \llbracket dx \rrbracket^2 \cdot n$ C/breathing, where the relative separation distance (dx) has been varied in the range of 2-20 µm.

**Table 1. TECD of pair of five selectie fabrics can be used in smart mask design**

| Pair | TSM ($+\sigma$) | TSM ($-\sigma$) | TECD (nC/m²) | | Effective TECD (nC/m²) |
|------|-----------------|-----------------|--------------|--|------------------------|
| 1 | Nylon | Polyester | -18.35(+) | -101.48 (-) | 83.13 |
| 2 | Cotton | Polyester | 110 (+) | -101.48 (-) | 211.48 |
| 3 | PMMA | PVDF | -48.73(+) | -87.35(-) | 38.62 |
| 4 | Nylon | PVDF | -18.35(+) | -87.35(-) | 69 |
| 5 | Polypropylene | Polyester | -27.23(+) | -101.48(-) | 74.25 |

**Figure 4**, depicting the variation voltage ($V$), current ($I$) and produced thermal power ($P$) induced per second in terms of the relative separation distance between two TLs. Each plot illuminates the variation $V$, $I$ and $P$ correspoding to the five different triboelectric fabric pairs. It is clearly indicating that the combination of nylon and polyester fabric found to be the best suited in this case.

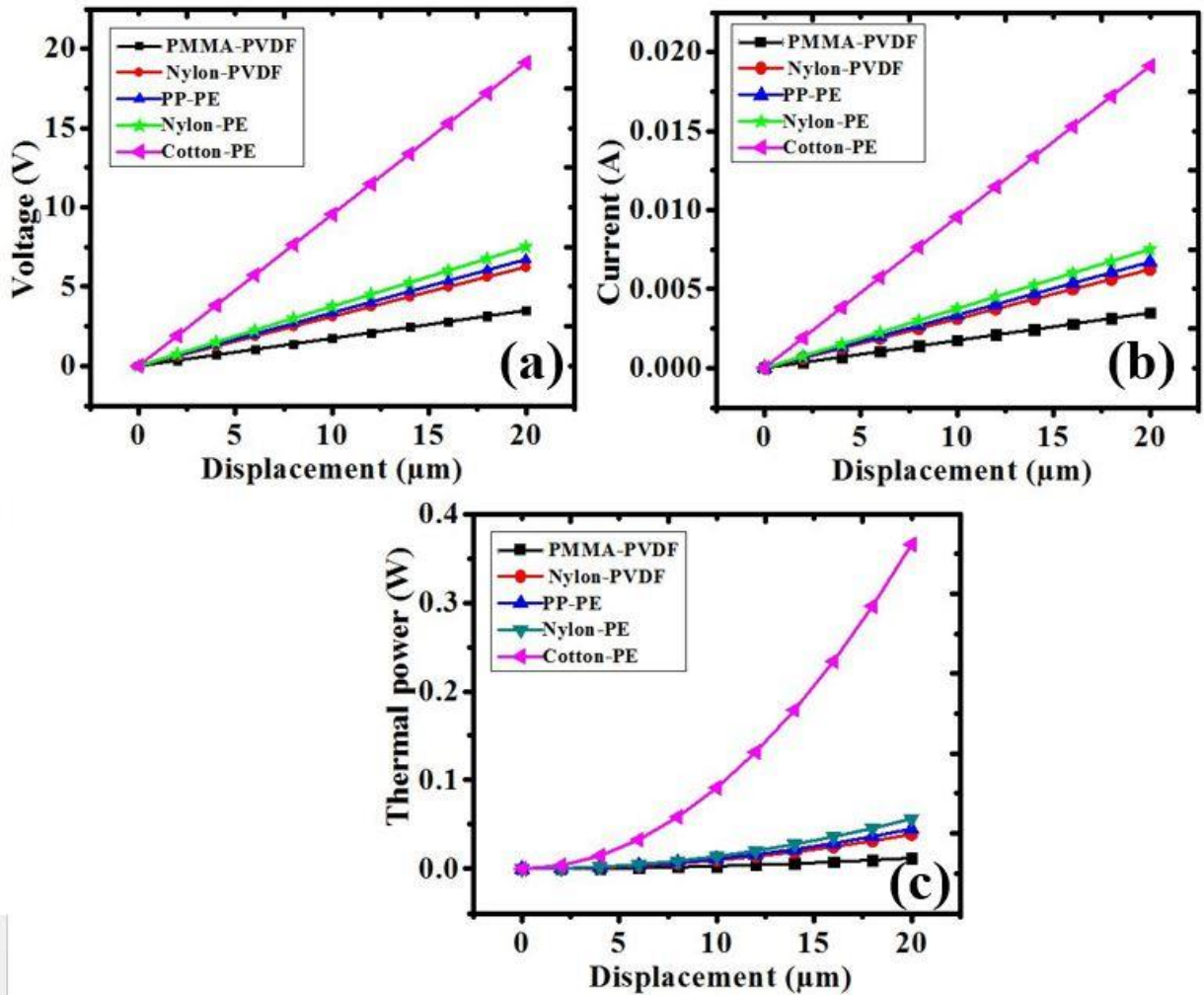

**Figure 4. Characteristics of voltage ($V$) vs. displacement ($dx$). current ($A$) vs. displacement ($dx$), thermal power ($W$) vs. displacement ($dx$).**

The proposed design explains that the outer smart layer, the tribo-charges induced through contact electrification and triboelectrification between tribo-pairs is likely to be utilized using the smart layer to kill the incoming SARS-CoV-2 contained aerosols. The storage capacitor ($C_s$) hold the induced charge in a way to be utilized in two different ways. It can serve effectively when the wearer is at rest and to retain the excess charge when incoming and outgoing aerosol contains no SARS-CoV-2. In addition, the incorporation of the smart outer layer provides extra protection to the wearer considering the case when incoming aerosol contained with SARS-CoV-2. In such cases, the charged SARS-CoV-2 can be killed by electrocution in the smart layer giving double protection through middle and inner TLs in self-powered way. In summary, the proposed technology can block the novel coronavirus through double action of charge adsorption, and electrocution by triboelectrification, thus providing effective protective role to combat the deadly impact of SARS-CoV-2. In light of this context, the reported masks based on different technologies have been illustrated in **Table 2**.

**Table 2. Comparison study of face masks**

| Type of mask | Type of fabric | Remarks |
|---|---|---|
| Telephone mouth piece mask[20] | Synthetic polymer (polyolyfin fiber) and electret treated non-woven web (meltbown web) | (i) It has designed to work using the principle of telephone handset.<br>(ii) The non-woven web is coated with a pressure sensitive adhesive so that the sound energy travels through the air into the microphone and makes the layer vibrate and respective layer converts the sound into electricity to make the outer layer electret so that incoming viral particle can be killed.<br>(iii) Repeatable usage of this mask can loosen the knitted threads thus chances of propagating sound wave get reduced. |
| Strapless flexible tribo-charged respiratory facial mask[21] | Multilayer flexible flat filter includes an activated carbon layer | (i)The filtration is based on the activation of carbon layer unlike triboelectricity, reusing the mask necessitated the refilling of carbon.<br>(ii) Working of such type of mask depends on the type of the skin of the wearer. The wearer might suffer from medical adhesive related skin injury (MARSI) |
| Multilayer composition for a breathing mask[22] | Internal and external spunbonded non-woven fabric, felt type tribocharged non-woven fabrics, a ply of melt-blown microfibre | (i) First intermediate layer of felt-type tribo-charged nonwoven fabric based on at least two differenttypes of fibres suitable for giving the fabric opposite electric charges that enhance the filtration.<br>(ii) The mask can filtrate particle sizes in the range of submicron. |
| Electrically charged filter and mask[23] | Four layered comprises of three layered liquid charged non-woven fibers and one layered tribocharged non-woven fabric. | (i) The induced temperature due to liquid charge intensity likely to b e less than $40^0$ C, which is insufficient to combat novel corona virus.<br>(ii) Refilling of polar liquid in the liquid charged fabric might be troublesome as it requires immersion apparatus alike spraying in the form of droplets, mist, shower etc. |
| Non-woven film and charged non-woven biological protection mask[24] | The inner layer made of rare earth material 'zein' and positive chitosan based outer layer. | (i) The mask basically deals with biological protection, and particularly relates to a nonwoven film and a charged non-woven biological protection mask.<br>(ii) A charged 'zein' based nanofiber double-layer film prepared through an electrospinning technique can isolate virus through the dual functions of electrical charge absorption and mechanical isolation. |
| Mask filter[25] | Sheet material composed of a resin fiber with wounding of copper wire. | (i) A copper wounded woven fiber sheet has been used for initiating corona discharge, i.e. the viral particles comes in close vicinity of the mask filter.<br>(ii) It is mentioned the mask is so designed to provide bactericidal effect unlike virucidal effect. |
| Mask using frictional and static electricity[26] | Polymer, nylon, cotton, silicon based polymer, polypropylene (PE), polypropylene terephthalate (PET) | (i) The design works based on the electrostatic and triboelectric properties.<br>(ii) The mask is its location specific, based on the country specific weather conditions (fine and yellow dust) the structure of the mask has been designed and it cannot be reusable. |
| Medical protective breathing mask[27] | The multylayer made of chitin fiber or silk fiber, hydrophilicandfwovenchemical fiber fabrics. | (i) The outgoing gas is transferred to the environment through the adsorption-diffusion-desorption process of the hydrophilic group of the functional film.<br>(ii) . The embodiment of the developed mask is pretty promising, but the working mechanism of such fiber including the contribution of chemicals involved here is quite ambiguous. |
| Respiratory protection mask[28] | Non-woven of melt-blown type fibers | (i) respiratory protection mask with greater breathability and to reduce breathing resistance.<br>(ii) The mask is intended to retain solid or liquid particles suspended in the air and in particular viruses or bacteria capable of causing diseases such as influenza. |
| Masks that use electrostatics of materials to protect healthy individuals from COVID 19[29] | Nylon cloth sandwiched between polypropylene layers | (i) The mask usable to adsorb viral particles between layers produced static electricity.<br>(ii) High chance to cross the electrostatic barrier as clinging on the surface of electrostatic layers requires a low pressure drop of incoming breathe. |
| Self-powered electrostatic adsorption face mask based on a tribo-electric nanogenerator[30] | Poly(vinylidene fluoride) electrospun nanofiber film (PVDF-ESNF) | (i) The ultrafine particulates are electrostatically adsorbed by the PVDF-ESNF, and the R-TENG can continually provide electrostatic charges in this adsorption process by respiration.<br>(ii) R-TENG, the SEA-FM shows that the removal efficiency of coarse and fineparticulates is higher than 99.20 wt. % and the removal efficiency of ultrafine particulates is still as high as 86.90 wt. % after continually wearing for 240 min. and a 30-day interval. |
| Washable Multilayer Triboelectric Air Filter for Efficient Particulate Matter PM2.5Removal[31] | Polytetrafluoroethylene (PTFE) and nylon fabrics | (i) It involves triboelectric air filter to filter out particulate matter (PM).<br>(ii) The mask is washable and exhibits removal efficiency of 84.7% for PM0.5, and 96.0% for PM2.5.<br>(iii) The whole filtration process is operated using linear motor to develop charge which increases the complexity of using the mask. |
| Aerosol Filtration Efficiency of Common Fabrics Used in Respiratory Cloth Masks[1] | Cotton−silk, cotton−chiffon, cotton−flannel | (i) Marks layer are particularly effective at excluding particles in the nanoscale regime (<∼100 nm),likely due to electrostatic effects that result in charge transfer with nanoscale aerosol particles.<br>(ii) The enhanced performance of the hybrids is likely due to the combined effect of mechanical and electrostatic-based filtration. |
| **Self-powered smart mask (This work)** | Nylon-polyester, cotton – polyester, PMMA –PVDF, nylon – PVDF, , polypropylene-polyester | (i) Design of the self-powered marks confirms the capability of the mask to function in response to talking, signing or any gestures of lips of the wearer with no difficulties of fetching external power source. <br> (ii) Tribo-series fabricsgenerates of static electricity and charged produced due static electricity further powers up the smart layer. <br> (iv) The smart layer is active during inhaling and exhaling period of wearing of the mask, which can be achieved by the capacitance connected with the electronic converter. <br> (v) Any virus-consist droplets/ aerosols can get electrified in the smart layer and furtherany virus get deactivated underthe tribo-field. <br> (vi) The proposed self-powered marks can generate thermal power in the range of 0.4W per second which is more than enough to electrocute virus-loaded aerosols. <br> (vii) Proposed mask is designed with simple elastic band for easy usage for every person including child. |

## 5. Conclusion

While the whole world has been suffering from the devastating COVID-19, safety precausion becomes a key concern to live life in the adverse situation. The present report illustrates the design of three-layered TENG based facial mask. The different combination of TSMs have been experimented in order to get better filtration efficiency in terms of TECD. The combination of easily available cotton and polyester fabric can be utilized in designing self-powered smart mask based on its highest TECD and induced power (0.38W per second). The study has also brought light into the voltage-current-power generated by the contact electrification of the TLs. The prototype mask can be activated through breathing cycles (and/or talking or other relevant facial gesture) without the need of any external power source. The accumulated charge can powered the smart layer upto 0.38 W which is sufficient enough to destroy the viral particles possibly by electrocution. The present design of facial mask can preferably reach the breakthrough in terms of cost-effectiveness, active-powered, double protection, and most importantly the wearer can reuse the mask throught their life span as it takes energy from the human body itself.

# Supplementary Information, Note S1

Note S1:

Work done by exhaling is given by

dw = P.dv (assuming pressure inside the plates to be constant)

and, dv = A.dx

Where, A = Active area.

dx = Relative displacement.

P = Atmospheric pressure.

If 'F' be the force exerted on the plate due to extra pressure,

Then,

dx= kdF (k is TSM dependent proportionality constant)

Considering,

- Volume of air/inhaling= 0.5L

- Breathing pressure= ±5 cm of $H_2O$ (water)

- Maximum pressure change during one breathing cycle =100 mm of water

- Normal air pressure=760 mm of Hg = 1.01325 bar

    Therefore, pressure variation during breathing is 1kPa.